\documentstyle[12pt]{article}
\begin{document}
\font\twmsbm=msbm10 scaled \magstep1
\font\temsbm=msbm10
\font\lamsbm=msbm10 scaled \magstep1
\def\@biblabel#1{\hbox{${}^#1$} \hfill}
\def\v#1{\hbox{\bf #1}}
\def\a{{\cal A}}
\def\z{\hbox{\twmsbm Z}}
\def\Z{\hbox{\lamsbm Z}}
\def\H{\hbox{\lamsbm H}}
\def\c{\hbox{\twmsbm C}}
\def\fn#1{${}^{\hbox{\scriptsize #1}}$}
\def\T{\hbox{\rm Tr}}
\def\tr{\T}
\begin{titlepage}
\begin{flushright}
TPJU - 7/92
\end{flushright}
\begin{center}
{\LARGE \bf Noncommutative Geometry
and Gauge Theory on Discrete Groups} \\ \ \\
{\em (revised version)} \\
\vspace{1in}
{\bf Andrzej Sitarz} \footnote{e-mail:
 ufsitarz@plkrcy11.bitnet} \\
\vspace{0.5cm}
{\sl Department of Field Theory} \\
{\sl Institute of Physics, Jagiellonian
University} \\
{\sl Reymonta 4, PL-30-059 Krak\'ow} \\
{\sl Poland}
\end{center}
\vfill
\begin{flushleft}
{\bf Abstract:\ }

We build and investigate a pure gauge theory
on arbitrary discrete groups. A systematic
approach to the construction of the
differential calculus is presented.  We
discuss the metric properties of the models
and introduce the action functionals for
unitary gauge theories.  A detailed analysis
of two simple models based on $\z_2$ and
$\z_3$ follows. Finally we study the method
of combining the discrete and continuous
geometry. \\
\end{flushleft}
\vfill
\end{titlepage}

\section{INTRODUCTION AND NOTATION}
\def\fn#1{\hbox{${}^{#1}$}}
The noncommutative geometry provides us
with a far more general
framework for physical theories than the
usual approaches. Its basic idea is to
substitute an abstract, associative and not
necessarily commutative algebra for the
algebra of functions on a smooth manifold
\fn{1-5}.  This allows us to use nontrivial
algebras as a geometrical setup for the
field-theoretical purposes, in particular
for the gauge theories, which are of special
interest both from mathematical and physical
points of view.

The construction of noncommutative gauge
theories has led to a remarkable result,
which is the description of the Higgs field
in terms of a gauge potential. This suggests
some possible nontrivial geometry behind the
structure of the Standard Model.  Several
examples of this kind, with various choices
of fundamental objects of the theory, have
been investigated in such context~\fn{6-13}.
The "discrete geometry" models, which take
as the algebra the set of functions on a
discrete space, seem to be one of the most
promising interpretations~\fn{4,5} and
suggest that such a geometry may play an
important role in physics.  Recently, some
more analysis has been carried out for two-
and three-point spaces~\fn{15,16} in the
context of grand unification and general
relativity.

We propose to develop here a systematic
approach towards the construction of a pure
gauge theory on arbitrary discrete groups.
The choice of a group as our base space is
crucial for our analysis and allows us to
make use of the correspondence with the
gauge theory on Lie groups.  The formalism
of finite derivations and invariant forms,
which we use in our approach, is equivalent
to the one used in various
works~\fn{2,4,12,13} for the two-point
space, it also extends considerably our
earlier studies~\fn{14}.

The paper is organized as follows: in the
first section we construct the tools of the
differential calculus, then we outline the
general formalism of gauge theories in this
case and some problems of the construction
of actions. The discussion of two examples
follows. Finally we discuss other
possibilities originating from the symmetry
principles and we present the method of
combining the discrete and continuous
geometry.

\section{DIFFERENTIAL CALCULUS}
\def\f{{\cal F}}
Let $G$ be a finite group and ${\a}$ be the
algebra of complex valued functions on $G$.
We will denote the group multiplication by
$\odot$ and the size of the group by $N_G$.
The right and left multiplications on $G$
induce natural automorphisms of $\a$, $R_g$
and $L_g$, respectively,
\begin{equation}
\left( R_h f \right)(g) \; = \; f(g \odot h),
\label{a0} \end{equation}
with a similar definition for $L_g$.

Now we will construct the extension of $\a$
into a graded differential algebra. We shall
follow the standard procedure of introducing
the differential calculus on manifolds, in
particular on Lie groups.  Therefore we use
almost the same terminology, although the
definitions of certain objects may differ.

First let us identify the vector fields over
$\a$ with linear operators on ${\a}$, which
have their kernel equal to the space of
constant functions. They form a subalgebra
of $GL(N_G,\c)$, with an additional
structure of a finite dimensional left
module over $\a$. Now, we can define the
vector space $\f$ of left invariant vector
fields as satisfying the following identity:
\begin{equation}
\partial \in \f \Leftrightarrow
\forall f \in \a \;\;\;\; L_h \partial(f) \; = \;
\partial( L_h f ).
\label{dd1}
\end{equation}

Before we discuss the algebraic structure of
$\f$ let us observe that this vector space
is $N_G-1$ dimensional and it generates the
module of vector fields.  This means that
for a given basis of $\f$, $\partial_i$,
$i=1\ldots N_G$, every vector field can be
expressed as a linear combination $f_i
\partial_i$, with the coefficients $f_i$
from the algebra ${\a}$.

${\f}$ forms an algebra itself and we find
the relations of generators to be of the
second order,
\begin{equation}
\partial_i \partial_j  \; = \;
\sum_k C_{ij}^k \partial_k, \label{a3}
\end{equation}
where $C_{ij}^k$ are the structure
constants. Because of the associativity of
the algebra they must obey the following set
of relations,
\begin{equation}
\sum_l C_{ij}^l C_{lk}^m \; = \; \sum_l C_{il}^m
C_{jk}^l. \label{a3a}
\end{equation}

Now we choose a specific basis of $\f$ and
calculate the relations (\ref{a3}) in this
particular case.  It is convenient for our
purposes to introduce the basis of ${\f}$
labeled by the elements of $G'=G \setminus
\{e\}$, where $e$ is the neutral element of
$G$. Further on, if not stated otherwise, it
should be assumed that all indices take
values in $G'$.
\begin{equation}
\partial_g f \; = \; f - R_g f, \;\;\; g \in G',\;
f \in {\a},
\label{dd2}
\end{equation}
The structure relations (\ref{a3}) become in
the chosen basis quite simple,
\begin{equation}
\partial_g \partial_h \; = \;
\partial_g + \partial_h -
\partial_{(h \odot g)}, \;\;\; g,h \in G'.
\label{da2}
\end{equation}
As a next step let us introduce the Haar
integral, which is a complex valued linear
functional on $\a$ that remains invariant
under the action of $R_g$,
\begin{equation}
\int f  \; = \; \frac{1}{N_G}
\sum_{g \in G} f(g), \label{a1}
\end{equation}
where we normalized it, so that $\int 1 =
1$.

Although the elements of $\f$ do not satisfy
the Leibniz rule, they are inverse to the
integration. Indeed, we notice that for
every $f \in \a$ and every $v \in \f$ the
integral (\ref{a1}) of $v(f)$ vanishes. For
this reason we can consider them as
corresponding to the derivations on the
algebra $\a$.

We define now the space of one-forms
$\Omega^1$ as a left module over $\a$, which
is dual to the space of vector fields. It
could be also considered as a right module
with an appropriate definition of the right
action of $\a$. This, however, is a
straightforward consequence of the
differential structure and we will discuss
it later.

Now we can introduce the notion of left
invariant forms, which, when acting on the
elements of $\f$, give constant functions.
Having chosen the basis of $\f$ we
automatically have the dual basis of ${\cal
F}^*$ consisting of forms $\chi^g$, $g \in
G'$, which satisfy
\begin{equation}
\chi^g(\partial_h)=\delta^g_h. \label{c1}
\end{equation}
To build a graded differential algebra we
need to construct n-forms and their products
for an arbitrary positive integer $n$. Of
course, we identify zero-forms with the
algebra $\a$ itself and their product with
the product in the algebra. The definition
for higher forms is natural, we take
$\Omega^n$ to be the tensor product of $n$
copies of $\Omega^1$,
\begin{equation} \Omega^n \; = \;
\underbrace{\Omega^1 \otimes \cdots
\otimes \Omega^1}_{\hbox{n
times}}, \label{dd3} \end{equation} and the
product of forms to be the tensor product
over $\a$.  However, let us remember that we
use here the tensor product of modules with
different right and left actions of $\a$.

To complete the construction of the
differential algebra we need to define the
external derivative $d$ and this is the
subject of the following lemma:

\begin{flushleft} {\bf Lemma}
\quad {\em There exists
exactly one linear operator $d$, $d:
\Omega^n \to \Omega^{n+1}$, which is
nilpotent, $d^2=0$, satisfies the graded
Leibniz rule and for every $f \in {\a}$ and
every vector field $v$ $df(v) = v(f)$,
provided that the right and left action of
$\a$ on ${\f}^*$ are related as follows,}
\end{flushleft}
\begin{equation}
     \chi^g f = (R_g f) \chi^g, \;\;\;\; g
\in G',\; f \in \a,
\label{dd4}
\end{equation}
\begin{flushleft}
{\em and that the following structure
relations hold,}
\end{flushleft}
\begin{equation}
d \chi^g  \; = \; - \sum_{h,k} C_{hk}^g
\chi^h \otimes \chi^k,
\;\;\;\; g \in G'.
\label{a5} \end{equation}

Before we prove the lemma, let us observe
that due to the properties of the tensor
product the condition (\ref{dd4}) could be
extended to the space of all one-forms.
Therefore it gives to $\Omega^1$ the
structure of a right module, mentioned
earlier. The next requirement (\ref{a5}) is
equivalent to the Maurer-Cartan structure
relations for Lie groups.

\noindent{\bf Proof: \quad }
Since we want $d$ to satisfy the
graded Leibniz rule, it is sufficient to
define the action of $d$ on $\a$ and on
${\f}^*$ because all other forms can be
represented as tensor products of them. The
action of $d$ on $\a$ is defined by the
requirement stated in the lemma, from which
we get that
\begin{equation}
 df \; = \; \sum_g (\partial_g f) \chi^g,
\label{dd5} \end{equation}
The Leibniz rule applied to the product of
any two elements $a,b
\in \a$, gives the following identity:
\begin{equation}
\sum_g \left( ab - R_g(a) R_g(b)
\right) \chi^g \; = \; \sum_g
\left(a
- R_g(a)\right) \chi^g b + a \left(b  -
R_g(b) \right) \chi^g,
\label{dd6} \end{equation}
which is satisfied only if (\ref{dd4})
holds.  The Maurer-Cartan relations arise
from the requirement that $d^2$ acting on an
arbitrary $a \in \a$ must vanish. Indeed, we
calculate,
\begin{equation} d^2 a \; = \; d
\left( \sum_h (\partial_h
a) \chi^h \right) \; = \end{equation} $$
\phantom{a} \; = \; \sum_{h,k} C_{hk}^g
(\partial_g a) \chi^h
\otimes \chi^k + \sum_h (\partial_h a) d \chi^h, $$
and this expression vanishes only if
(\ref{a5}) is true.  This ends the proof.

In our construction we have obtained the
differential algebra over the algebra of
complex functions on a discrete group, which
may be a starting point for the analysis of
this structure. One may attempt, for
instance, to calculate its cohomology. Let
us notice that although the basic algebra
was commutative, in the end we obtained a
noncommutative, infinite-dimensional
algebra, which may be an interesting subject
of further studies in the program of
non-commutative geometry.  However, in this
paper we shall rather proceed towards the
construction of gauge theories on the basis
of introduced formalism.

Let us end this section by constructing the
involution on our differential algebra,
which agrees with the complex conjugation on
$\a$ and (graded) commutes with $d$, i.e.
$d(\omega^\star) = (-1)^{\hbox{\scriptsize
deg} \omega} (d
\omega)^\star$.
Again, it is sufficient to calculate it for
one-forms,
\begin{equation}
(\chi^g)^\star \; = \; - \chi^{g^{-1}}.
\label{dd7}
\end{equation}

So far, we restricted ourselves in our
approach to the complex-valued functions.
Similarly we can consider a straightforward
extension of the model if we take functions
valued in any involutive algebra, for
instance, the matrix valued functions. The
quotient subalgebras of the obtained algebra
may also be considered, the necessary
formalism and the examples will be given in
the last section.

\section{GAUGE THEORY}
\def\h{{\cal H}}
\def\aa{\tilde{\cal A}}
\subsection{General Formalism}
In this section we shall construct the gauge
theory on finite groups using the
differential calculus we have just
introduced.  First let us explain some basic
ideas. The starting point is the
differential algebra $\tilde{\Omega}^*$ with
its subalgebra of zero-forms $\aa \subset
\tilde{\Omega}^*$. We take the group of
gauge transformations to be any proper group
${\h} \subset \aa$, which generates $\aa$.
In particular, we will often take $\h$ to be
the group of unitary elements of $\aa$, $$\h
= {\cal U}(\aa) = \{a \in \aa: \; aa^\star =
a^\star a = 1
\}.$$

Of course, the external derivative $d$ is
not covariant with respect to the gauge
transformations. Therefore we have to
introduce the covariant derivative $d+\Phi$,
where $\Phi$ is a one-form.  The requirement
that $d+\Phi$ is gauge covariant under gauge
transformations,
\begin{equation}
d + \Phi  \; \rightarrow \; H^{-1} (d +
\Phi) H,
\;\;\;\; H \in \h, \label{b0}
\end{equation}
results in the following transformation rule
of $\Phi$,
\begin{equation}
\Phi \; \rightarrow \; H^{-1}
\Phi H + H^{-1} d H. \label{b1}
\end{equation}
$\Phi$ is the gauge potential, which we will
also call connection. If the gauge group is
unitary, we require also that the covariant
derivative is hermitian,
\begin{equation}
(d+\Phi) \left( a^\star b \right) \; = \;
a^\star (d+\Phi) b + \left(b^\star (d+\Phi)
a \right)^\star,
\;\;\; a,b \in \aa,
\end{equation}
which results in the condition that the
connection is anti-selfadjoint,
$\Phi=-\Phi^\star$.  Finally, we have the
curvature two-form, $F = d\Phi + \Phi \Phi$,
which, of course, is gauge covariant.

In order to proceed with the construction
and analysis of the Yang-Mills theory we
have to introduce a metric.  We shall
briefly mention here a general theory and
concentrate our efforts in the next section
on the analysis of the model under study.

Let us define the metric $\eta$ as a form on
the left module of one-forms, valued in the
algebra $\aa$ and bilinear over the algebra
$\aa$,
\begin{equation}
\eta:\;\; \tilde{\Omega}^1 \times
\tilde{\Omega}^1 \; \mapsto \;
\aa,
\label{e19}
\end{equation}
$$ \eta(av,ub) \; = \; a \eta(v,u) b, \;\;\;
a,b \in \aa, \; u,v
\in
\tilde{\Omega}^1.$$
Note that $\eta$ can no longer be symmetric.
One can consider, however, the $\c$-valued
bilinear functional on $\tilde{\Omega}^1$,
which is the composition of the metric
$\eta$ and the integral on $\aa$. The latter
can be any \c-linear functional on $\aa$,
which is symmetric, gauge invariant and
real-valued on self-adjoint elements of
$\aa$, if the algebra is involutive. Then
the new functional can be made symmetric
provided that we impose some restrictions on
$\eta$. We shall discuss it in details for
our particular model. If we additionally
require that $\eta(\omega,
\omega^\star)$ is self-adjoint
for an involutive $\aa$, we
immediately notice that the composition may
be used to construct a semi-norm on
$\Omega^1$

For the purpose of this paper and the
studies of discrete geometry the above
definition of the metric is sufficient,
although in the case of more complicated
non-commutative algebras a more detailed
analysis would be necessary. However, this
shall be the task of future investigations,
here we restrict the detailed study to the
main subject of discrete geometry. Let us
only say that in general case one needs to
extend the metric $g$ to the vector spaces
of forms of higher order as well as to
introduce the already mentioned integral.

\subsection{Gauge Transformations,
Connection and
Curvature on Discrete Groups}
\def\ab{{\hbox{\rm A}}}
\def\hb{{\hbox{\rm H}}}

Let us take the algebra $\aa$ to be the
tensor product of the algebra $\a$ of
complex valued functions on $G$, which we
introduced in the previous section, by a
certain algebra $\ab$, which could be the
algebra of complex $n \times n$ matrices
$M_n$, for instance. In such case, the
differential algebra $\tilde{\Omega}^*$ is
clearly the tensor product of $\Omega^*$ by
$\ab$. The group of gauge transformations,
as defined above, can be identified with the
group of functions on $G$ taking values in a
group $\hb \subset \ab$. Similarly, gauge
potentials are interpreted as $\ab$ valued
one-forms. We will denote the involution on
$\ab$ by $\dagger$.

Due to this simplified structure it is
sufficient to construct the metric only for
the differential algebra $\Omega^*$, as
described in the previous section, since it
could be extended immediately for the whole
algebra. A natural interpretation of this
property is that the metric shall depend
only on the base space of the gauge theory,
characterized by $\a$, and not on the target
space, characterized by $\ab$. We take the
integral to be the combination of the Harr
integral, as defined in (\ref{a1}) and the
trace operation on the algebra $\ab$, which
we assume to exist.

Before we introduce the metric, let us work
out the gauge transformation rules
(\ref{b1}) for the connection and the
curvature in the convenient basis we chose
(\ref{c1}).  If we write $\Phi = \sum_{g}
\Phi_g \chi^g$, the transformation of
$\Phi_g$ under gauge transformation $H \in
\h$ is
\begin{equation} \Phi_g \;
\rightarrow \; H^{-1} \Phi_g
(R_g H) + H^{-1} \partial_g H \label{b2}
\end{equation}

The condition $\Phi = - \Phi^\star$ enforces
the following relation of its coefficients,
\begin{equation}
\Phi^\dagger_g \; = \; R_g \left( \Phi_{g^{-1}} \right),
\label{b2y}
\end{equation}
If we introduce a new field $\Psi = 1 -
\Phi$, $\Psi_g = 1 -
\Phi_g$,
we can see that (\ref{b2}) is equivalent to
\begin{equation}
\Psi_g  \; \rightarrow \;
H^{-1} \Psi_g  (R_g H). \label{b2a}
\end{equation}
The introduction of $\Psi$ is convenient for
the calculations as it simplifies the
formulas. We will discuss the physical
meaning of this step later. It is
instructive to compare formulas for the
coefficients of the curvature, $$F =
\sum_{g,h} F_{gh} \chi^g \otimes \chi^h,$$
using both $\Phi$ and $\Psi$. In the first
case, from the definition of $F$, the rules
of differential calculus
(\ref{dd4},\ref{a5}) and the exact form of
the structure constants in this basis
(\ref{dd2}), we obtain,
\begin{equation}
F_{gh} \; = \; \Phi_{(h \odot g)} - \Phi_g -
R_g (\Phi_h) +
\Phi_g R_g(\Phi_h),
\label{b3b} \end{equation}
whereas the same formula written with $\Psi$
is much simpler,
\begin{equation}
F_{gh} \; = \;  \Psi_g R_g(\Psi_h) -
\Psi_{(h \odot g)}.
\label{b3bb}
\end{equation}
The transformation rule for $F_{gh}$ follows
from the gauge covariance of $F$. However,
since the algebra is noncommutative the
coefficients are no longer gauge covariant:
\begin{equation}
F_{gh} \; \rightarrow \; H^{-1} F_{gh}
(R_{(h \odot g)} H).
\label{b3a}
\end{equation}

\subsection{The Yang-Mills Action}
In our investigation of gauge theories in
the setup of discrete geometry we have come
to the point when we need to introduce the
action.  Therefore we shall now discuss the
problem of the metric.  Suppose, we have a
metric $\eta$ defined on the space of
one-forms $\Omega^1$.  By $\eta^{gh}\in\a$
we denote its values on the elements of the
basis, $\eta^{gh} = \eta(\chi^g,\chi^h)$.
Clearly, this metric is not symmetric,
nevertheless, if we require that after the
integration we should recover the symmetry,
\begin{equation}
\int \eta(u,v) \; = \; \int \eta(v,u),
\label{sy} \end{equation}
we obtain from the definition of the Haar
integral (\ref{a1}) and the relations
(\ref{dd4}) that this is possible only if
$\eta^{gh}$ are \c-numbers such that
$\eta^{gh} = \eta^{hg} \sim
\delta^{gh^{-1}}$.
Now let us consider $\ab$-valued one-forms.
Since we are dealing with the gauge theory
let us postulate the most natural
requirement in such case, which is that the
metric is gauge covariant, i.e.,
\begin{equation}
\int \tr \; \eta(u,v) \; = \;
\int \tr \; \eta(H^{-1}uH, H^{-1}vH),
\;\;\; H \in {\cal H}, \; u,v \in \aa.
\label{sy2}
\end{equation}
The bilinearity of $\eta$, as defined
previously, gives us immediately,
$$\eta(H^{-1}uH, H^{-1}vH) = H^{-1}
\eta(uH,H^{-1}v) H, $$
so after taking the trace we need only to
check the last part of the equality,
\begin{equation}
\eta(uH,H^{-1}v)
= \sum_{g,h} u R_g(H) \eta^{gh}
R_{h^{-1}}(H^{-1}) v \; =
\label{sy3} \end{equation}
$$ \; = \;
\sum_{g,h} \eta^{gh} u_g \left( R_g(H)
R_{h^{-1}}(H^{-1}) \right) v_h. $$

The right-hand side of the last equality is
gauge invariant provided that $\eta^{gh}
\sim \delta^{gh^{-1}}$, which is again the
condition obtained earlier by requiring the
symmetry of the integrated metric. In
addition, if we want $\eta(u,u^\star)$ to be
self-adjoint, we must fix $\eta^{gh}$ to be
real numbers.

Now, we can tackle the analogous problem of
the metric structure on the space of
two-forms. This would allow us to construct
the Yang-Mills action. Following the
arguments above, for any two forms $u,v \in
\Omega^2$, which have a unique
representation as $u=\sum_{g,h} u_{gh}
\chi^g \otimes \chi^h$ and $v=\sum_{g,h} \chi^g
\otimes \chi^h v_{gh}$, let us construct
the bilinear form,
\begin{equation}
\theta(u,v) \; = \; \sum_{g,h,g',h'} u_{gh}
\theta^{ghg'h'} v_{g'h'}, \label{y0} \end{equation}
where each $\theta^{ghg'h'}$ is in the
beginning an arbitrary element of $\a$. The
bilinearity is again mixed, i.e. from the
left for the first entry and from the right
for the other one. The elements
$\theta^{ghg'h'}$ are the evaluation of the
metric on the basis of two-forms,
$\theta(\chi^g \otimes \chi^h,
\chi^{g'} \otimes \chi^{h'})$.

Again, we require that after integration the
metric must be symmetric and that it remains
gauge invariant. This leads to the
restriction that $\theta^{ghg'h'}$ has to be
a \c-number, which vanishes unless $h' \odot
g' = g^{-1} \odot h^{-1}$ and additionally,
$\theta^{ghg'h'} = \theta^{g'h'gh}$.  Since
we want to construct $\theta$ from the
metric $\eta$ we obtain, after taking into
account the conditions above, the following
general expression,
\begin{equation}
\theta^{ghg'h'} \; = \; \alpha \eta^{gh} \eta^{g'h'}
+ \beta \eta^{gh'} \eta^{g'h},
\label{y1} \end{equation}
where $\alpha,\beta$ are arbitrary
constants.

In the case of a commutative group $G$ we
have also an additional term possible,
\begin{equation}
\theta^{ghg'h'}_{c} \; = \;
\theta^{ghg'h'} + \gamma \eta^{gg'} \eta^{hh'}.
\label{y4} \end{equation}

We want the Yang-Mills action to be
constructed in the same way as in the case
of the gauge theories on manifolds.
Therefore, we postulate that for an
involutive algebra and the structure group
$\hb \subset {\cal U}(\ab)$, it has the
following form:
\begin{equation}
S_{\hbox{\tiny YM}} \; = \; \int \tr \;
\theta( F, F^\star),
\label{YM}
\end{equation}
where $\int$ is the Haar integral on $\a$
and $\tr$ is the trace on $\ab$. First of
all, from the previous considerations we
immediately notice that (\ref{YM}) is indeed
gauge invariant. This formula, together with
corresponding expressions for the metric
$\theta$ applied in the situation of
calculus on manifolds yields the standard
answer. For noninvolutive algebras or other
structure groups one has to modify the
expression for the 'squared norm' of $F$,
which we used in the formula (\ref{YM}).

Finally let us calculate the action
(\ref{YM}) using the functions $F_{gh}$.
Let us denote by $\eta^{gh^\star}$ the value
of $\eta(\chi^g,(\chi^h)^\star)$. From the
definition of the metric (\ref{y0},\ref{y1})
and from the involution rules on our algebra
(\ref{dd7}) we get,
\begin{equation}
S_{\hbox{\tiny YM}}=
\int \sum_{g,h,g',h'} \left(
\alpha \eta^{gh} \eta^{g'^\star h'^\star}  \tr \, F_{gh}
F_{h'g'}^\dagger + \beta \eta^{gh'^\star}
\eta^{hg'^\star}
\tr \, F_{gh} F_{h'g'}^\dagger \right).
\label{y5}
\end{equation}

Of course, for the commutative group $G$ we
get also another term coming from
(\ref{y4}).  Therefore, due to the arbitrary
choice of the constants $\alpha,\beta$, we
can single out at least two possible
independent actions, which are of the second
order in $F$.  If we take into account the
form of the metric,
\begin{equation}
\eta^{gh} \; = \; E_g \delta^{gh^{-1}}, \label{YM0}
\end{equation}
together with the involution rule
(\ref{dd7}) we obtain the following
expressions:
\begin{eqnarray}
S_{1} \; &=& \; \int \sum_{g,h} E_g
E_{h^{-1}} \tr \; F_{gg^{-1}}
F^\dagger_{hh^{-1}}, \label{YM1} \\ S_{2} \;
&=& \; \int \sum_{g,h} E_g E_{h^-1}
\tr \; F_{gh} F_{gh}^\dagger, \label{YM2}
\end{eqnarray}
and again, for commutative $G$, we
additionally get,
\begin{equation}
S_{c} \; = \; \int \sum_{g,h} E_g E_h
\tr \; F_{gh} F_{hg}^\dagger. \label{YM3}
\end{equation}

All these actions are of course gauge
invariant and independent of our choice of
the basis, the latter due to their
construction, which involves the contraction
of tensors (\ref{YM}).  The properties of
$\eta$ guarantee also that they are all
real.

Let us point out that the constructed
actions or rather each possible linear
combination of them may pretend to be the
Yang-Mills action of our theory. This
ambiguity is the result of the fact that our
differential algebra is noncommutative,
furthermore it is interesting that it also
depends on the group structure of the base
space.  Another significant feature of the
theory is that the space of possible metrics
on $\Omega^*$ is much smaller than one would
expect.

Finally, let us point out that one could
build more invariant quantities, which might
be used in the construction of the general
action of our gauge theory.  For instance,
let us notice that the following quantity,
\begin{equation}
S_m \; = \; \int \tr \;
\sum_{g,h} \eta^{gh} F_{gh},
\label{YM4}
\end{equation}
is gauge invariant and independent of our
choice of the basis. If we rewrite it using
the shifted connection $\Psi$, use the exact
form of $\eta^{gh}$ (\ref{YM0}) and take
into account (\ref{b2y}) we obtain
\begin{equation}
S_m \; = \; \int \tr \;
\sum_{g} E_g \Psi_g \Psi_{g}^\dagger,
\label{YM4a}
\end{equation}
which is of course real. So, this term is as
good as all introduced earlier and therefore
it also has to be taken into account.

In the above analysis we used the metric
$\eta$ to construct the actions.  Let us now
discuss the issue of other possible
candidates, which can replace the metric.
For instance, we can introduce another
bilinear functional on $\Omega^1$, which is
this time bilinear with respect to the left
action of $\a$ on its both entries,
$\rho(au,bv)=ab\rho(u,v)$.  This holds only
for commutative $\a$ and can be extended for
$\h$ valued forms after taking the trace.
Additionally, $\rho$ can be made symmetric
and gauge invariant if $\rho^{gh} =
\rho(\chi^g,\chi^h) \sim \delta^{gh}$. The
only missing property is that
$\rho(a,a^\star)$ may not be self-adjoint
and therefore the actions may appear not to
be real valued.

We constructed the theory in a purely
geometric way. Thus, although we chose our
specific basis, which proved to be very
convenient for the calculations, our results
are independent of this choice.  We shall
not discuss this symmetry here, let us only
mention that if we neglect it we end up with
more candidates for gauge invariant objects.

\section{EXAMPLES}
In this section we will study two simple
examples of the unitary gauge theory on the
two- and  three-point spaces, with the
structure of abelian groups.  We will
construct the action functionals and discuss
briefly the solutions and their geometry.
\def\si{\hat{\Psi}}
\subsection{Gauge Theory on $\Z_2$}

Let us denote the group elements of $\z_2$
by $+$ and $-$.  We take the group $\hb$ to
be $U(N)$ and the algebra $\ab$ to be the
algebra of complex matrices $M_n$.  Since
the structure group is unitary, the
connection must be antihermitian, therefore
we obtain the following relation,
\begin{equation}
\Phi_-(+) \; = \; \Phi_-^\dagger(-), \label{b5}
\end{equation}
and the same applies to $\Psi_- = 1 -
\Phi_-$. Thus, effectively we have got only
one degree of freedom, which is an arbitrary
complex matrix. We take it as $\si =
\Psi_-(+)$ and for convenience we drop here
the subscript index.

Let us observe that for $n \geq 1$ all
$\Omega^n$ are one-dimensional.
Consequently, the curvature two-form $F =
F_{--} \chi^- \otimes \chi^{-}$ is
completely determined by one coefficient
function $F_{--}$, which using
Eqs.~(\ref{b3bb}) and (\ref{b5}) can be
calculated to be,
\begin{eqnarray}
F_{--}(+) & = &  \si \si^\dagger -1,
\label{b7a} \\ F_{--}(-) & = &
\si^\dagger \si -1. \label{b7b} \end{eqnarray}

Of course the metric is trivial in this
case, for simplicity we take $\eta^{--} =
1$. Now one can easily see that all
possibilities for the Yang-Mills action are
reduced to the following,
\begin{equation}
S_{YM} \; = \;  \tr \; ( \si^\dagger \si - 1
)^2,
\label{b8}
\end{equation}
where we have already done the Haar
integration.

This has exactly the form of the potential
of the Higgs model and was first obtained in
Connes' consideration of the $\c^2$
algebra~\fn{1}. Here, however, we can modify
this expression slightly by adding the term
linear in $F$ (\ref{YM4}), which is
proportional to $\tr \; (\si\si^\dagger
-1)$.  In this case we get the total action
equivalent to (\ref{b8}) with the field
$\Psi$ rescaled.

Let us now make some comments on the moduli
space of the theory and the extremal points
of the action functionals. The space of flat
connections modulo gauge transformations is
trivial. Indeed, the vanishing of $F$ is
equivalent to the unitarity of $\si$ and
from its transformation rule (\ref{b2a}) we
see that arbitrary $\si$ can be obtained
from the trivial flat connection, $\Psi =
1$, by choosing the appropriate gauge
transformation.  The Yang-Mills action has
one absolute minimum, which is reached for
the flat connections.

\def\p{\partial}
\subsection{Gauge Theory on $\z_3$}

Let us denote the elements of the group by
$0,1,-1$ and the group action by $+$. The
space of one-forms is two-dimensional,
spanned by the basis of invariant forms
$\chi^+,\chi^-$, (for the indices, $+$
stands for $+1$ and $-$ for $-1$).

We choose the metric $\eta$ to be in its
simplest form, so that
$\eta^{+-}=\eta^{-+}=1$ and the other two
components vanish according to our
requirements (\ref{YM0}).

The algebra of derivations $\p_+,\p_-$ obeys
the following relations,
\begin{eqnarray}
\p_- \p_- & = & 2 \p_- - \p_+, \label{c1a} \\
\p_+ \p_+ & = & 2 \p_+ - \p_-, \label{c1b} \\
\p_- \p_+ & = & \p_+ \p_-  \; = \;
\p_+ + \p_- , \label{c1c}
\end{eqnarray}
which determine the structure constants and
the rules of the differential calculus in
this case (\ref{a5}).

Now, let us again construct the $U(N)$ gauge
theory. The gauge potential one-form $\Phi$
could be expressed as $\Phi_+ \chi_+ +
\Phi_- \chi_-$. The condition that $\Phi$ is
antihermitian (\ref{b2y}) takes the form,
\begin{equation}
\Phi_+(g) \; = \; \Phi_-^\dagger(g+1),
\;\;\; g \in \z_3.  \label{c3}
\end{equation}
{}From this relation we see that the
connection is completely determined by
either of its coefficients. Let us define
$\Psi = 1 - \Phi_+$, and use it in further
analysis. Its gauge transformation is as
follows,
\begin{equation}
\Psi(x) \; \rightarrow  H^\dagger(g) \Psi(g)
H(g+1). \;\;\; H(g) \in U(N), \; g \in \z_3.
\label{c4}
\end{equation}
Now we can express the curvature in terms of
the function $\Psi$.  Since $\Phi_+ = 1-
\Psi$ and $\Phi_- = R_-(1 - \Psi^\dagger)$
we obtain the coefficients $F_{gh}$,
\begin{eqnarray} F_{++} & = &  \Psi
(R_+ \Psi) - R_- \Psi^\dagger, \label{c5a}
\\ F_{+-} & = &  \Psi (\Psi^\dagger) - 1,
\label{c5b} \\ F_{-+} & = &  (R_-
\Psi^\dagger) (R_- \Psi) - 1, \label{c5c} \\
F_{--} & = &  (R_- \Psi^\dagger)
(R_+\Psi^\dagger) - \Psi. \label{c5d}
\end{eqnarray} One can easily notice that
$F_{--} = R_+ F_{++}^\dagger$ and both
$F_{+-}$ and $F_{-+}$ are hermitian.

Before we discuss the action functionals let
us find the moduli space of flat connections
in this case. If $F$ vanishes, from
(\ref{c5b}) we obtain that the function
$\Psi$ must be unitary, whereas $F_{++}=0$
gives us from (\ref{c5a}) and from the
previous result the following identity,
\begin{equation}
(R_- \Psi) \Psi (R_+ \Psi) \; = \; 1.
\label{c6} \end{equation} Using the
transformation rule (\ref{c4}) and the
condition above we can again show that all
flat connections are gauge equivalent.

Finally, let us present the actions. The
action linear in $F$ is,
\begin{equation}
S_m  \; = \;  2 \int \tr \; \left( \Psi
\Psi^\dagger -1 \right),
\label{c7}
\end{equation}
and we are left with three possible terms
        for the Yang-Mills quartic action,
\begin{eqnarray}
S_1 &=& 2 \int \tr \; \left( (\Psi
\Psi^\dagger - 1) R_- (\Psi^\dagger \Psi_
-1) + (\Psi \Psi^\dagger-1)(\Psi^\dagger
\Psi -1) \right),
\label{c8} \\
S_2 &=& \int 2 \tr \; \left(
(\Psi \Psi^\dagger -1)^2
+ R_+ (\Psi \Psi^\dagger)
(\Psi^\dagger \Psi) \right. \nonumber \\
\; &+& \left. \Psi^\dagger \Psi -
R_-(\Psi) \Psi R_+(\Psi) -
R_+(\Psi^\dagger) \Psi^\dagger
R_-(\Psi^\dagger) \right), \label{c8a} \\
S_{c} &=&
2 \int \tr \; \left(
(\Psi \Psi^\dagger - 1) R_- (\Psi^\dagger \Psi_ -1)
+ R_+ (\Psi \Psi^\dagger)
(\Psi^\dagger \Psi) \right. \nonumber \\
\; &+& \left. \Psi^\dagger \Psi -
R_-(\Psi) \Psi R_+(\Psi) -
R_+(\Psi^\dagger) \Psi^\dagger
R_-(\Psi^\dagger) \right) \label{c8b}
\end{eqnarray}
Each linear combination of them may pretend
to be the global action of the theory. Let
us observe the remarkable property that
there exists a combination, which is a third
order polynomial in $\Psi$,
\begin{equation}
S_{3}  \; = \; 2 S_{1} + 2 S_{m} - S_{1} -
S_{m} \; = \; 2 \int \tr \; R_-(\Psi)
\Psi R_+(\Psi) + \hbox{c.c.}
\label{c10}
\end{equation}
Such decomposition of quartic actions into
the third-order invariants is not a general
feature of the theory and is characteristic
only for the model under study.

The problem of the extremal points of the
presented actions is more complicated that
in the previous example and in some cases
the action might not even have an absolute
minimum.  The analysis becomes much simpler
if we restrict ourselves to the
consideration of smaller algebras, which we
shall discuss in the next section.

\section{SYMMETRIES AND SUBALGEBRAS}

This section is devoted to a brief
discussion of possible restrictions and
modifications of the theory, which arise
from employing symmetries of the considered
algebra.
\def\bb{\tilde{\cal B}}
Suppose we take a proper subalgebra of
$\aa$, we shall denote it by $\bb \in \aa$.
Then we can find a graded differential
subalgebra of $\tilde{\Omega}$ in such a way
that the zero-forms are the elements of
$\bb$ and all the rules of differential
calculus remain unchanged. For instance, let
us consider the subalgebra of $\a$, which
contains all \c-valued constant functions on
the group $G$ and denote it by $\a_0$. If we
take as $\Omega^n$ all differential forms
that have their coefficients in $\a_0$, we
obtain the required subalgebra of
$\Omega^*$. We can express that construction
in a more formal way. Indeed, if we use the
group of automorphisms of $\a$, $R_g, g \in
G$, which can be easily extended to the
whole of $\Omega^*$, we see that
$\Omega_0^*$ remains invariant under the
action of this group.  Therefore, this
construction can be generalized for any
algebra and any group of its automorphisms.
A trivial example from the differential
calculus on Lie groups is set by the
restriction to the algebra of left-invariant
forms. In the particular case of the $SU(2)$
gauge theory constructed in such model was
discussed in details in several
papers~\fn{6-10}, although it was not
formulated precisely in the language we
introduced here. More sophisticated model,
with the coset symmetry condition, was
analysed using this formalism~\fn{17}.

Let us turn our attention back to the
introduced examples. Having the differential
grading algebra $\Omega^*_0$ we can proceed
with the construction of gauge theories
following the steps we have already made for
$\Omega^*$.

The gauge transformations are now global,
i.e. they are also constant, when considered
as functions on $G$. The same applies to the
coefficients of the connection and the
curvature. Let us see what is the effect of
this for the constructed theories on $\z_2$
and $\z_3$.

In the first case, from (\ref{b5}) we get
that $\Psi$ must be hermitian. This implies
that the minima of the Yang-Mills action
(\ref{b8}), which again correspond to the
flat connections, are separated. The moduli
space is equal to the space of equivalence
classes of unitary and hermitian matrices.
Notice that for $U(1)$ we obtain $\z_2$,
which is the base space of our theory $G$.

The same applies to the model with $\z_3$.
This time, the moduli space of flat
connections is the space of equivalence
classes of matrices that satisfy the
relation $\Psi^3=1$. Again, for $U(1)$ this
space can be identified with $\z_3$.  The
arbitrariness in the choice of action is
reduced slightly when compared to the model
with larger algebra, as in this case we have
$S_2 = S_m$.

We shall end here the discussion of
possibilities arising from employing the
symmetry principles encoded in the
automorphisms of the algebras. Our aim was
only to show this option and present briefly
the implications for the given models and we
leave a more detailed investigation of this
topic for future studies.

\section{PRODUCTS OF DISCRETE AND CONTINUOUS GEOMETRY}

Suppose we have two graded differential
algebras, $\Omega_1,\Omega_2$, with the
external derivative operators $d_1$ and
$d_2$ respectively. Then we can construct
the tensor product of them, which can be
made again a differential algebra provided
that we take,
\def\ot{\tilde{\otimes}}
\begin{equation}
d ( \omega_1 \ot \omega_2) \; = \; (d_1
\omega_1) \ot \omega_2 +
(-1)^{\hbox{\scriptsize deg}
\omega_2} \omega_1 \otimes (d_2 \omega_2),
\label{pp1}
\end{equation}
Let us take as the first component the
algebra of differential forms on a manifold
and  the differential algebra over a
discrete group as the other one. Their
tensor product is the basis for the
description of models combining both
continuous and discrete geometry. Remember
that we have to distinguish the tensor
product of these algebras, $\ot$ from the
product in $\Omega_2$, which we denote again
by $\otimes$ and the product within
$\Omega_1$, denoted as usually by $\wedge$.

We shall calculate here, as a very
illustrative example, the outcome of a
unitary gauge theory on two base spaces.
The first will be the product of the
Euclidean space $M$ and the two-point space
and the second will arise from the same
euclidean geometry and from the algebra of
constant functions on the three-point space
$\z_3$.

In the first example, we may assume simply
that the algebra of zero-forms consists of
complex functions, with their arguments from
$M \times \z_2$. The gauge group is then
$U(1)$.  We know the action of the external
derivative on each separate algebra as well
as the rule (\ref{pp1}). Thus, the gauge
connection $\Phi$ is a one-form, comprising
discrete and continuous geometry
differential forms:
\def\dm#1{{\partial #1 \over \partial x^\mu}}
\def\dn#1{{\partial #1 \over \partial x^\nu}}
\def\dmu#1{{\partial #1 \over \partial x_\mu}}
\begin{equation}
\sum_\mu A_\mu dx^\mu + {\Phi} \chi^-.
\label{pp2}
\end{equation}
Both  $A_\mu$ and $\Phi_-$ are again
functions on the product space,
$A_\mu=A_\mu(x,g)$ and $\Phi=\Phi(x,g) =
R_-(\Phi(x,g))^*$.  Let us calculate the
curvature, which in addition to the
'continuous and discrete' terms has the
mixed ones as well.  We use the terminology
from the section 2.1, with the shifted
connection $\Psi$. Notice that because
$\dm{\phantom{a}}$ and $ \partial_-$
commute, the product of forms dual to them
anticommutes: $dx^\mu \ot \chi^- = - \chi^-
\ot dx^\mu$.  After some calculations we
finally get,
\begin{eqnarray}
F & \; = \; & \sum_{\mu,\nu}
\left(\dm{A_\nu} - \dn{A_\mu} \right) dx^\mu
\wedge dx^\nu \nonumber +(\Psi \Psi^* - 1)
\chi^- \otimes \chi^-  \\ & + & \left(
\dm{\Psi} -  A_\mu \Psi +  R_-(A_\mu) \Psi
\right) dx^\mu \ot \chi^-
\label{pp3}
\end{eqnarray}

If we take the metric to be the tensor
product of metrics on each algebra and
integrate over $M$ with respect to the
standard measure, we obtain the following
Yang-Mills action,
\begin{eqnarray}
S_{\hbox{\scriptsize YM}} = &
\int_M   F_{\mu\nu}(x,+) F^{\mu\nu}(x,+) +
F_{\mu\nu}(x,-) F^{\mu\nu}(x,-) &  \nonumber
\\ & + (\Psi(x,+) \Psi^*(x,+) - 1)^2 &
\label{pp4} \\ & + \left( D_\mu \Psi(x,+)
\right)
\left( D^\mu \Psi(x,+) \right)^*, & \nonumber
\end{eqnarray}
where we have used properties of the metric
and integration on the two-point space and
the relation $A_\mu = - A_\mu^*$.  $D_\mu$
is the abbreviation, which we use for the
following operator:
\begin{equation}
D_\mu \Psi \; = \; \dm{\phantom{a}} -
A_\mu(x,+) + A_\mu(x,-), \label{pp4a}
\end{equation}
The action (\ref{pp4}) describes a scalar
$U(1)\times \bar{U}(1)$ Higgs model, with a
scalar complex field $\Psi(x,+)$ and two
gauge fields $A_\mu(x,+)$ and $A_\mu(x,-)$.
The interaction term for the scalar field,
which has the form of the quartic potential,
arises in this way naturally. Let us point
out the importance of the fact that the
physical meaning is given not to the gauge
potential $\Phi$ but to the shifted
connection $\Psi$. Only by using $\Psi$ we
get the coupling between discrete and
continuous parts in the form of the
covariant derivative (\ref{pp4a}).

Similarly as above we will now briefly
discuss the unitary gauge theory on the
product of continuous geometry on $M$ by the
geometry set by the invariant subalgebra
$\Omega_0$ over $\z_3$. Here we shall not
restrict ourselves to the $U(1)$ case, as
nonabelian case would be more interesting.
So we have the gauge connection in the
product algebra,
\begin{equation}
A_\mu(x) dx^\mu + \Phi(x) \chi^- +
\Phi(x)^\dagger \chi^+,
\label{pp5}
\end{equation}
and the curvature two-form {\bf F},
\begin{eqnarray}
\hbox{\bf F} & = & \left( \dm{A_\nu} - \dn{A_\nu}  +
\left[ A_\mu , A_\nu \right] \right)
dx^\mu \wedge dx^\nu + \sum_{g,h = +,-}
F_{gh} \chi^g \otimes
\chi^h
\nonumber \\
& + & \left( \dm{\Psi} + \left[ A_\mu , \Psi
\right]  \right) dx^\mu \ot \chi^ - - \left(
\dm{\Psi^\dagger} + \left[ A_\mu ,
\Psi^\dagger \right]
\right) dx^\mu \ot \chi^ +,
\end{eqnarray}
where again $\Psi=1-\Phi$ and one has to
insert the exact form of $F_{gh}$ from
(\ref{c5a}-\ref{c5d}). We observe that due
to the underlying $\z_3$ space we have some
freedom in the choice of our action.  It is
clear that it must include the following
terms,
\begin{equation} \tr \; F_{\mu\nu}^\dagger F^{\mu\nu} +
\tr \;  ( D_\mu \Psi ) ( D_\mu \Psi)^\dagger
\end{equation}
where $D_\mu$ is the covariant derivative
$D_\mu \Psi = \dm{\Psi} + [ A_\mu, \Psi]$.
The ambiguity arises when we consider the
self-interaction term for $\Psi$. We can
choose any of the possible actions
(\ref{c8}-\ref{c8b}), which could give the
same quartic potential as in the previous
example but we can get the $\Psi^3$ type
action from (\ref{c10}) as well.

Finally let us notice that if the gauge
group is abelian we have no coupling between
the $A$ and $\Psi$ fields.

\section{CONCLUSIONS}
We presented in this paper a systematic
approach to the problem of construction of
the gauge theories on discrete spaces. This
involved the introduction of the
differential calculus, which we have carried
out for spaces that possess the structure of
a finite group.  This structure was a key
point of our analysis, so we established
that in order to construct theories on
discrete spaces one has to specify their
group properties as well. It would be
interesting, for instance, to investigate
theories, which have the same number of
points in the base space but which differ in
their group structures.

We constructed gauge theories only for the
unitary gauge groups, indicating that the
formalism could be extended to arbitrary
groups.  In fact, they do not have to be
continuous and one may as well use discrete
groups for this purpose. Another spectacular
property of this theory is the fact that we
can take as a starting point the algebra,
which is not necessarily the algebra of
\c-valued functions or functions valued in
any algebra but its proper subalgebra.  For
instance, if have an algebra $A$, and its
subalgebras, say $A_0,A_1,\ldots$, we can
construct the proper subalgebra of $\aa$ as
the set of functions such that $f(x), x \in
G$ takes values in $A_j$ for some index $j$.
Now, following the same steps as we
presented in this paper, we can construct
the differential calculus and the gauge
theories. It appears that if we consider the
two-point space, as in the first example,
with the algebra of functions taking values
in {\c} at one point and at the other in
{\H}, which is the algebra of quaternions,
its product with the continuous geometry of
the Minkowski space gives us the precise
description of the pure gauge part of the
electroweak interactions. Of course, in this
approach fermions stay out of the picture.

The discussion of the Yang-Mills action has
led us to the investigation of the metric
properties of the model. This seems to be
another interesting point arising from our
considerations and in the future
investigations~\fn{18} we shall attempt a
thorough discussion of this topic.
Consequently, one may try to investigate the
general relativity of discrete geometries
alone or of the product of them by the
continuous manifolds, which may have a deep
physical meaning if the geometry of the
world involves discrete part, as suggested
by the Standard Model.

Finally, let us notice that the restriction
to the finite groups may be relaxed as well
and one can analyze similar models for
infinite discrete groups like $\z_n$, for
example.

The program of noncommutative geometry has
given us the possibility of considering a
far more general class of models than the
one arising from the analysis on manifolds.
Since quantum groups and discrete geometry
are the two most interesting and promising
examples, their study seems to be important
and we believe that their analysis, in
particular the investigation of gauge
theories in this framework, will help to a
better understanding of the subject.

\vfill
\pagebreak[4]
\def\v#1{{\bf #1}}

\vfill
\end{document}